\documentstyle[12pt,epsfig]{article}
\def\be{\begin{equation}} \def\ee{\end{equation}} \def\bea{\begin{eqnarray}} 
\def\eea{\end{eqnarray}} \def\nnb{\nonumber} 
\def\gat{\tilde{g}_A}
\begin{document}
\renewcommand{\thefootnote}{\fnsymbol{footnote}}
\setcounter{footnote}{1}

%\hfill{June 3, 2004 \ \ \ {\tt Lnbeta30}}
\hfill{TRI-PP-04-05,\ \ USC(NT)-04-04}

\begin{center}
\vskip 5mm 
{\Large\bf 
Neutron beta decay in effective field theory}
\vskip 5mm 
{\large 
S. Ando$^{(a,c)}$\footnote{E-mail:sando@triumf.ca},
H.~W. Fearing$^{(a)}$\footnote{E-mail:fearing@triumf.ca},
V. Gudkov$^{(c)}$\footnote{E-mail:gudkov@sc.edu},
K. Kubodera$^{(c)}$\footnote{E-mail:kubodera@sc.edu},\\
F. Myhrer$^{(c)}$\footnote{E-mail:myhrer@physics.sc.edu},
S. Nakamura$^{(b,c)}$\footnote{E-mail:nakamura@yukawa.kyoto-u.ac.jp},
T. Sato$^{(b,c)}$\footnote{E-mail:tsato@phys.sci.osaka-u.ac.jp} 
} 
\vskip 5mm

{\it 
${}^{(a)}$Theory Group, TRIUMF, 4004 Wesbrook Mall, 
Vancouver, B.C. V6T 2A3, Canada \\
${}^{(b)}$Department of Physics, Osaka University, 
Toyonaka, Osaka 560-0043, Japan \\
${}^{(c)}$Department of Physics and Astronomy,
University of South Carolina,
Columbia, \\
SC 29208, USA}

\end{center}

\vspace{8mm}

Radiative corrections to the lifetime 
and angular correlation coefficients 
of neutron beta-decay 
are evaluated in 
effective field theory. 
We also evaluate 
the lowest order nucleon recoil corrections, 
including weak-magnetism. 
Our results agree with 
those of the long-range and model-independent
part of previous calculations.  
In an effective theory the model-dependent 
radiative corrections are replaced by 
well-defined low-energy constants.  
The effective field theory allows a
systematic evaluation of 
higher order corrections to our results 
to the extent that 
the relevant low-energy constants are known. 

\vskip 5mm \noindent
PACS: 23.40.-s, 13.40.Ks, 12.39.Fe, 11.30.Rd

\newpage
\renewcommand{\thefootnote}{\arabic{footnote}}
\setcounter{footnote}{0}
\noindent {\bf 1. Introduction  } 

The radiative corrections for beta decay 
have been intensively investigated 
by a number of authors,
and the prime issue for such studies 
has been to deduce the value of the
Cabbibo-Kobayashi-Maskawa (CKM) 
matrix element $V_{ud}$ from 
nuclear beta-decay data.
An accurate 
value for $V_{ud}$ is important 
for testing the unitarity of the CKM matrix. 
The most precise values of $V_{ud}$ have been 
obtained from the accurate data of 
super-allowed $0^+\to 0^+$ 
nuclear beta-decays~\cite{Vud}.
Neutron beta-decay measurements provide 
an alternative method of determining  
$V_{ud}$, a method which does 
not depend on the accuracy of nuclear models.
Neutron beta-decay experiments 
also provide the most precise 
determination of the 
axial-vector coupling constant, $g_A$, 
which plays an important role 
in hadronic weak-interaction reactions
including many astrophysical processes. 
Theoretically, pion beta-decay can also be used 
for determining $V_{ud}$.
Unfortunately, however,
the currently available experimental data on pion beta-decay 
are not accurate enough to allow us to take 
full advantage of this merit, 
see {\it e.g.} Cirigliano {\it et al.}~\cite{cirigliano03}.

To extract an accurate value of $V_{ud}$ 
from neutron decay data,
the theoretical expression 
for the neutron decay rate 
including radiative corrections 
must be known with sufficient accuracy.
The usual convention is to decompose radiative 
corrections of order $\alpha$ into two parts, 
the ``outer" and the ``inner" 
corrections~\cite{sirlin,sirlinrmp,marciano-sirlin}. 
The ``outer" correction is 
a universal function of the electron energy, 
independent of the details of the strong interactions. 
The ``inner" correction stems from 
short-range terms and
hadronic structure effects. 
This hadronic structure dependence 
(and additional nuclear structure dependence
in the case of nuclear beta-decay)
causes uncertainties in extracting 
fundamental quantities like $V_{ud}$
from experimental data \footnote{
A new calculation of these
radiative corrections, 
obtained with the standard model of electroweak
interactions, has been reported in a recent
preprint~\protect{\cite{bunatian}}. 
The results however seem to differ markedly
from the classic calculations of Sirlin {\it et al.}
\protect{\cite{sirlin,sirlinrmp,marciano-sirlin}}.
}.

In this communication
we present the first calculation 
of radiative corrections to 
neutron beta-decay based on  
a low-energy effective field theory (EFT).  
EFT provides symmetry constraints
required by the underlying theory and 
a systematic expansion scheme 
for the evaluation of the hadron current. 
As suggested by Weinberg~\cite{weinberg},  
low-energy hadronic physics can be described by an 
effective field theory of QCD 
known as ``chiral perturbation theory'' ($\chi$PT).
The effective chiral Lagrangian, ${\cal L}_\chi$,
reflects the symmetries 
and the pattern of symmetry breaking 
of the underlying QCD. 
For massless quarks the QCD Lagrangian is 
chirally symmetric, 
but chiral symmetry is spontaneously broken
generating the pions as massless Goldstone bosons.
Since the $u$ and $d$ quark masses are very small
compared with the QCD scale $\Lambda_{\rm QCD}$,
and since the finite pion mass
generated by the quark masses
is small compared to 
a typical strong interaction scale,
it is reasonable to treat the explicit 
chiral symmetry breaking terms 
as small perturbations. 
${\cal L}_\chi$ is expanded in powers of
$Q/\Lambda_\chi \ll 1$ 
where $Q$ denotes the typical four-momentum  
of the process in question or 
the pion mass, $m_\pi$, which represents the small explicit 
chiral symmetry breaking scale. 
The chiral scale,  
$\Lambda_\chi \simeq 4\pi f_\pi \simeq $ 1 GeV 
($f_\pi = 92.4$ MeV is the pion decay constant),   
is associated with the ``high-energy" processes  
that have been integrated out in arriving at ${\cal L}_\chi$
and with pion loops.
The parameters appearing in ${\cal L}_\chi$,
called the {\it low-energy constants} (LEC's),
effectively subsume the high-energy physics 
that has been integrated out.
In principle, these LEC's could be determined 
from the underlying theory,   
but in practice the LEC's are 
determined phenomenologically from 
experimental data. 
Once the LEC's are determined from appropriate 
empirical data, 
then ${\cal L}_\chi$ represents a complete 
Lagrangian up to a 
specified chiral order.
Furthermore, starting from ${\cal L}_\chi$,
one can develop, for the amplitude 
of a given process,
a well-defined perturbation scheme
by organizing  the relevant Feynman diagrams 
according to powers in 
$Q/\Lambda_\chi$. 
If all the Feynman diagrams up to a given power, $\nu$, 
in $Q/\Lambda_\chi$ are taken into account, 
then the results depend only on 
the LEC's up to this order, 
with the contributions of higher order terms 
suppressed by an extra power of $Q/\Lambda_\chi$.

Over the past decade 
$\chi$PT has been successfully applied 
to many processes; for reviews, see, 
{\it e.g.}, Refs.~\cite{bkm95,scherer}. 
Chiral Lagrangians including the photon field 
have been developed and applied to, {\it e.g.}, 
pion-nucleon scattering, see Refs.~\cite{muller,fettes2}.
Our present calculation of the radiative corrections 
to neutron beta-decay 
is an EFT based on the spirit of the
chiral Lagrangian approach.
Thus we write down an effective Lagrangian,
appropriate to neutron beta-decay,
obeying chiral symmetry and involving 
a minimum set of LEC's 
and use the Lagrangian to 
estimate the relevant amplitudes 
to leading, next-to-leading, 
and next-to-next-to leading orders
(LO, NLO, N$^2$LO) in the
$Q/\Lambda_\chi$ expansion.
In fact, since the typical energy transfer of the reaction
is much smaller than the pion mass,
the ``$Q/\Lambda_\chi$ expansion" here has a special feature
to be explained in the next section.

The results of our EFT calculation confirm 
the expression for 
the model-independent universal function 
derived by Sirlin~\cite{sirlin}. 
Furthermore, our calculation provides 
expressions for corrections of order $\alpha$
to the angular correlation coefficients 
in neutron beta-decay. 
We will show that the short-distance phenomena including  
the model-dependent hadronic radiative corrections 
can be condensed into two LEC's, 
one relevant to the Fermi constant $G_F$ and 
the other to the axial coupling constant $g_A$.  
The values of these LEC's 
need to be determined by experiments.  
In order to have crude order-of-magnitude 
estimates of our LEC's,  
we also compare our results with
the ``inner'' radiative corrections obtained in
the standard calculations. 
Furthermore, we shall argue that,
provided the LEC's involved in our calculation
are of a ``natural" size,  
the neutron-decay rate and angular correlation 
coefficients calculated here
are expected to have a precision 
better than $10^{-3}$.  

\vspace{3mm} \noindent 
{\bf 2. Effective theory for neutron beta-decay } 

Since neutron beta-decay is a low energy process,
it is natural to use here 
heavy-baryon chiral perturbation theory (HB$\chi$PT), 
see, {\it e.g.}, Refs.~\cite{bkm95,scherer}. 
In fact the appropriate amplitude,
however without radiative corrections, 
can be obtained from HB$\chi$PT
calculations of muon capture on a proton,
$\mu+p\to n+\nu$,
which have been carried out including N$^2$LO correction 
terms~\cite{fear97,bern98,ando00,bern01}. 
Neutron beta-decay, however, has a feature not
shared by muon capture, namely several different expansion scales. 
In particular, the 
maximum energy release, 
$\Delta M =m_n-m_p-m_e=0.782$ MeV,
is very small compared to the pion mass $m_\pi$ 
and the nucleon mass $m_N= (m_p+m_n)/2$. 
Correspondingly, if we denote by $\bar{Q}$
the typical four-momentum transfer of the process,
$\bar{Q}\sim \Delta M$ 
is also very small. 
We therefore introduce here 
a particular ``$Q/\Lambda_\chi$" expansion
in which $Q$,
unlike most HB$\chi$PT calculations,
only represents $\bar{Q}$.
The chiral symmetry breaking scale, 
$m_\pi/\Lambda_\chi \simeq 0.14$, 
will be accounted for separately.  
The nucleon recoil terms are governed by 
the scale $\bar{Q}/m_N \simeq 0.8 \times 10^{-3}$,  
and they are NLO corrections to the LO expression. 
The scale $\bar{Q}/m_N \simeq \bar{Q}/\Lambda_\chi$ is 
numerically of the same 
magnitude as $\alpha/(2\pi) \sim 10^{-3}$,  
governing the radiative corrections, 
which are our primary
interest ($\alpha $ is the fine structure constant). 
Therefore,
for our present purposes,
we consider the 
$\alpha/(2\pi)$ and $\bar{Q}/m_N$ corrections 
to be of the same order. 

The relevant effective Lagrangian, ${\cal L}_\beta$,
for the neutron decay process reads
\bea
{\cal L}_\beta &=& 
  {\cal L}_{e\nu\gamma} 
+ {\cal L}_{NN\gamma} 
+ {\cal L}_{e\nu NN},
\label{eq;Lag}
\eea
where ${\cal L}_{e\nu\gamma}$ is 
the lepton-photon Lagrangian,  
${\cal L}_{NN\gamma}$ describes the heavy nucleon 
interacting with a photon,
and ${\cal L}_{e\nu NN}$ gives the effective 
$V-A$ interaction
between the lepton and the heavy nucleon current.
Since the pion mass is much heavier than the typical 
momentum scale of the reaction, $\bar{Q}\ll m_\pi$,
we suppress the pion fields of
the chiral Lagrangian, ${\cal L}_\chi$, 
and in ${\cal L}_\beta$
we have retained only the interactions 
between the heavy nucleon field, 
lepton current, and photons.
Later in the text, 
we will discuss the role of the pions
in the present calculation.
Thus one has, through LO and NLO,
\bea
\lefteqn{{\cal L}_{e\nu\gamma} = 
- \frac{1}{4} F^{\mu \nu}F_{\mu \nu} 
-\frac{1}{2\xi_A}(\partial\cdot A)^2
+\left(1+\frac{\alpha}{4\pi}e_1\right)
\bar{\psi}_e \; (i\gamma\cdot D )\psi_e 
-m_e\bar{\psi}_e \psi_e 
+\bar{\psi}_\nu i\gamma \cdot \partial \psi_\nu , }
\\
\lefteqn{{\cal L}_{NN\gamma} = 
\bar{N}\left[
1+\frac{\alpha }{8\pi}e_2(1+\tau_3) 
\right]
iv\cdot D N ,}
\\
\lefteqn{{\cal L}_{e\nu NN} = 
- \frac{( \stackrel{\circ}{G}_F V_{ud} )}{\sqrt{2}}\,
\bar{\psi}_e\gamma_\mu(1-\gamma_5)\psi_\nu 
\left\{
\bar{N}\tau^+
\left[ 
\left(1 + \frac{\alpha}{4\pi} e_V \right)v^\mu
-2 \stackrel{\circ}{g}_A \left(1 + \frac{\alpha}{4\pi} e_A 
\right)S^\mu
\right]N 
\right. }  
\nnb \\ 
&&+ \left. \frac{1}{2m_N}\bar{N}
\tau^+\left[ 
i(v^\mu v^\nu -g^{\mu\nu})
(\stackrel{\leftarrow}{\partial}-\stackrel{\to}{\partial})_\nu
-2i \stackrel{\circ}{\mu}_V
[S^\mu,S\cdot(\stackrel{\leftarrow}{\partial}
 +\stackrel{\to}{\partial})]
- 2 i\stackrel{\circ}{g}_A v^\mu S\cdot 
(\stackrel{\leftarrow}{\partial}-\stackrel{\to}{\partial})
\right] N \right\} ,
\nnb \\ 
\label{eq;V-A}
\eea
where $F_{\mu\nu}=\partial_\mu A_\nu-\partial_\nu A_\mu$ and 
$D_\mu$ is the covariant derivative of QED.
The $\xi_A$ is the gauge parameter 
and we choose the Feynman gauge $\xi_A=1$.
The $v^\mu$ is 
the velocity vector of 
the heavy-baryon formalism,
which we take as $v^\mu=(1,\vec{0})$,
and $S^\mu$ is the nucleon spin operator 
$2S^\mu = (0,\vec{\sigma})$. 
The isovector magnetic moment in 
the NLO Lagrangian is 
$\stackrel{\circ}{\mu}_V \to \mu_V=4.706$.
The quantities $e_1$, $e_2$, $e_V$ and $e_A$ are 
defined as the LEC's of the theory.
The LEC's $e_1$ and $e_2$ are the $\alpha$-order corrections
related to the wave-function normalization factors 
of the electron and proton, respectively.
The LEC's $e_V$ and $e_A$ are the $\alpha$-order corrections
to the Fermi and Gamow-Teller amplitudes,
where we have factored out the common coefficient 
$(\stackrel{\circ}{G}_FV_{ud})/\sqrt{2}$.
Those LEC's are used to absorb infinities coming from 
the virtual photon-loops and 
take into account short-range radiative effects.
We remark that some of those LEC's contain  
contributions from, 
{\it e.g.}, $g_i$'s for the one nucleon sector
without leptons\cite{muller} 
and $X_i$'s for 
the meson sector with leptons\cite{knecht} in $\chi$PT
\footnote{Unfortunately the connection between 
the LEC's $e_V$ and $e_A$ and 
the $g_i$ and $X_i$ of 
\protect{Refs.~\cite{muller,knecht}} is not straightforward. 
The $g_i$ and $X_i$ originate in Lagrangians
involving only subsets of the degrees of freedom considered here 
and thus generate radiative corrections to only particular vertices in the
diagrams for neutron beta decay. Their contribution can be absorbed in
$e_V$ and $e_A$, but $e_V,e_A$ would also contain contributions from the
LEC's of a yet-to-be-calculated Lagrangian involving 
the nucleon, lepton current, and photons simultaneously.}.
As is conventional, the parameters of the initial Lagrangian, {\it e.g.} 
the Fermi constant 
$\stackrel{\circ}{G}_F$ 
and the axial coupling constant 
$\stackrel{\circ}{g}_A$,   
are taken as the coupling constants in the absence 
of radiative corrections and
in the chiral limit, $m_\pi = 0$. 
Thus in particular, 
we assume that the Fermi constant,
$\stackrel{\circ}{G}_F \to G_F$ = 1.166 
$\times 10^{-5}$ GeV$^{-2}$, 
as determined from muon-decay.
As we discuss in the next paragraph, 
higher order hadronic corrections, 
{\it i.e.}, pion-loops,
renormalize these ``bare" couplings
to their physical values in the absence of electromagnetic effects, 
{\it e.g.} 
$\stackrel{\circ}{g}_A \to g_A$. 
Furthermore, radiative effects give rise to 
additional corrections to the coupling constants,
$G_F$ and $g_A$ which depend on the process being considered. 
These radiative corrections 
will be displayed explicitly in the present work. 

\begin{figure}
\begin{center}
\epsfig{file=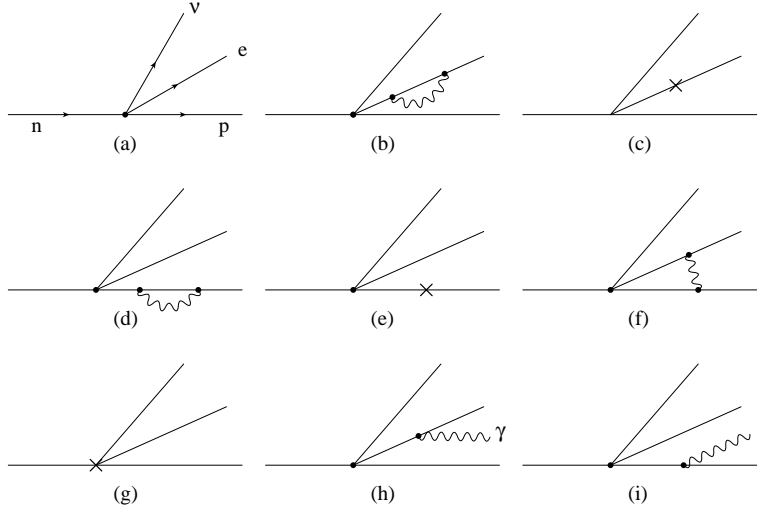,width=10.0cm}
\caption{
Feynman diagrams for neutron beta-decay 
up to order $\alpha$.  
In diagram (a), the four-fermion vertex
can represent either the leading-order (LO) 
or next-to leading order (NLO) vertex,
the latter being a $1/m_N$ correction to the former. 
The crosses on the electron and nucleon lines 
in diagrams (c) and (e) are 
vertices involving the LEC's, $e_1$ and $e_2$, respectively. 
The vertex of diagram (g) is given by 
the LEC's $e_V$ and $e_A$. 
}
\label{fig;diagrams}
\end{center}
\end{figure}

We calculate the Feynman diagrams 
shown in Fig. \ref{fig;diagrams}, 
where the vertices are determined by 
the Lagrangian, ${\cal L}_\beta$, given above. 
Several remarks are in order on the diagrams
in Fig. \ref{fig;diagrams}.
Consider first diagram (a),
which does {\em not} involve radiative corrections.  
Diagram (a) is a 
tree-diagram for the LO and NLO amplitudes.
As regards the LO contribution, 
one may wonder why we do not consider here
the pion-pole diagram (not shown). 
The pion-pole diagram, which is responsible 
for the induced pseudoscalar coupling, 
formally belongs to LO and hence 
would be included in normal circumstances. 
However, the extremely small momentum transfer 
involved in neutron beta-decay ($\bar{Q} \ll m_\pi$) 
drastically suppresses the pion-pole diagram contribution.
Due to the presence of the pion propagator
and a momentum of order $\bar{Q}$ at each vertex, 
the pion-pole diagram scales like
$(\bar{Q}/m_\pi)^2\simeq 3 \times 10^{-5}$
relative to the dominant LO terms. 
The accuracy of our present treatment 
does not warrant the inclusion of this
tiny pion-pole contribution,
and we will not consider it 
in the main body of our calculation. 
In the concluding section, however,
we will briefly discuss 
the LO pion-pole term and its radiative corrections. 
Diagram (a) in Fig.~\ref{fig;diagrams} 
also includes the NLO vertex
coming from the nucleon recoil terms 
$\propto \bar{Q}/m_N$ featuring in Eq.~(\ref{eq;V-A}). 
Since we are treating 
the $\bar{Q}/m_N$ and $\alpha/(2\pi)$ corrections 
as contributions of the same order,  
we will discuss these recoil terms later in the text; 
however, in evaluating radiative corrections,
we need not consider the recoil terms 
since these corrections would be of 
higher order 
$\sim \alpha/(2\pi)\times \mu_V \bar{Q}/(2 m_N) 
\sim 10^{-6}$.  
At order N$^2$LO there occur two kinds of contributions. 
Higher order recoil corrections scale 
as $(\bar{Q}/m_N)^2\simeq 10^{-6}$ 
and therefore can be neglected. 
The remaining N$^2$LO terms (diagrams not shown) 
come from pion-loops and 
the corresponding hadronic LEC's
which would appear in HB$\chi$PT Lagrangian at this order, 
see {\it e.g.} Refs.~\cite{bkm95,scherer}. 
The pion-loop diagrams which generate terms proportional 
to $\bar{Q}^2$ 
{\it i.e.}, terms representing
the hadronic vertex form factor effects,
can be neglected,
since their contributions are suppressed
by a factor of $(\bar{Q}/\Lambda_\chi)^2 \simeq 10^{-6} $
relative to the dominant LO terms. 
The remaining contributions of the pion-loops, 
which contain terms proportional to 
$(m_{\pi}/\Lambda_\chi)^2$, 
renormalize the bare quantities such as 
the ``bare'' axial vector coupling constant 
$\stackrel{\circ}{g}_A$.  
These $(m_\pi/\Lambda_\chi)^2$ terms and the corresponding 
hadronic LEC's 
are absorbed into the renormalized $g_A$ so that 
to N$^2$LO order, 
$g_A = \; \stackrel{\circ}{g}_A 
[ 1+ {\cal O}( (m_\pi/\Lambda_\chi)^2) ] $, 
see {\it e.g.} Eq.~(4.50) in Ref.~\cite{bkm95} 
or Eq.~(50) in Ref.~\cite{fear97}. 
Radiative corrections to the pion loop diagrams 
are suppressed by a scale 
$(m_\pi/\Lambda_\chi)^2 \simeq 2 \times 10^{-2}$ 
relative to the 
leading \underline{radiative} corrections, 
and therefore their contributions can
be ignored in the present calculation.

The above discussion indicates that,
to the accuracy in question,
we need only consider radiative corrections
of the following type.  
Of the contributions topologically represented by diagram (a),
consider those involving the LO vertex
and evaluate all possible radiative corrections applied
to these LO diagrams.
Diagrams (b), (d), (f) in Fig.~\ref{fig;diagrams} 
are one-photon loop corrections 
for the electron propagator, 
the nucleon propagator, 
and the four-point vertex function, respectively.
Meanwhile, diagrams (c), (e) and (g)
represent the contributions of the counter terms,
the $e_1$, $e_2$, $e_V$ and $e_A$ terms,
in the Lagrangian.
These LEC's remove the
ultraviolet divergence arising from 
the loop diagrams (b), (d) and (f).
As is well known,
the infrared divergences contained 
in diagrams (b), (d), (f)
should be canceled by the infrared divergences 
in the bremsstrahlung diagrams (h) and (i)\footnote{
Recently these bremsstrahlung diagrams have been studied 
by Bernard {\it et al.} for radiative neutron beta decay,
$n\to p+\nu+e+\gamma$, 
in EFT\cite{bgmzRNBD}.
},
and we have confirmed this cancellation 
explicitly. 

\vskip 3mm \noindent
{\bf 3. The correlation coefficients and the decay rate from EFT. } 

A general expression 
for the differential neutron decay rate
d$\Gamma$ is well known \cite{jtw57}
for a case wherein only the neutron is polarized,
and in which the nucleon recoil and 
radiative corrections are ignored: 
\bea
\frac{d\Gamma}{dE_ed\Omega_{\hat{p}_e}d\Omega_{\hat{p}_\nu}}
&\simeq& 
\frac{(G_FV_{ud})^2}{(2\pi)^5}(1+3g_A^2)
|\vec{p}_e|E_eE_\nu^2 
\left[
1 + a\;  ({\vec{\beta}\cdot\hat{p}_\nu} ) 
+ b \left(\frac{m_e}{E_e}\right) 
\right.  \nonumber \\ &&  \left.
+ \hat{n}\cdot \left( 
  A \; {\vec{\beta}} 
+ B\; {\hat{p}_\nu} 
+D \; \frac{\vec{p}_e\times \vec{p}_\nu}{E_eE_\nu}
\right) \right] .
\label{eq;aAB}
\eea 
Here $E_e$ and $\vec{p}_e$ ($E_\nu$ and $\vec{p}_\nu$) are the
electron (neutrino) energy and momentum, 
$\hat{n}$ is the neutron spin
polarization vector, $\vec{\beta}=\vec{p}_e/E_e$,  
and $a,b,A,B,D$ are the correlation coefficients.
If we calculate   
diagram (a) in Fig.\ref{fig;diagrams}
in the LO approximation, 
and if we neglect the nucleon recoil terms
in the phase space factor, 
then our calculation reproduces 
Eq. (\ref{eq;aAB}), and furthermore 
we recover the standard lowest order expressions 
for the correlation coefficients as given in \cite{jtw57}:
\bea
a = \frac{1-g_A^2}{1+3g_A^2},
\ \ \ 
A = \frac{-2g_A^2+2g_A}{1+3g_A^2},
\ \ \ 
B = \frac{2g_A^2+2g_A}{1+3g_A^2}, 
\label{eq;gA}
\eea
where $g_A$ is the physical axial coupling constant.
The coefficient $b$ in Eq.~(\ref{eq;aAB}),
which reflects the presence of
scalar and tensor weak couplings,
vanishes in our LO calculation,
since our Lagrangian only contains 
the standard vector and axial vector weak
interaction.  
The parameter $D$ in Eq.~(\ref{eq;aAB}) is related to
time-odd correlations and hence
it also should vanish in the LO calculation
since our Lagrangian is $T$ invariant. 
However, ``induced" $D$ terms can 
appear at higher orders.
For instance, interference 
between the weak magnetism and the radiative corrections
would generate a $D$ term
of order $10^{-5}$\cite{ct-pr67}.

As we proceed to include the higher order radiative diagrams 
generated by the
Lagrangian of Eq.~(\ref{eq;Lag}), we encounter 
infinities coming from the photon-loop diagrams 
in Fig.~\ref{fig;diagrams}. 
In order to eliminate these infinities,
we need to introduce counter terms with the
corresponding LEC's in our Lagrangian. 
We renormalize these LEC's 
in the usual effective field theoretical method 
based on the
dimensional regularization of 
loop integrals~\cite{bkm95,scherer}.
The finite LEC's renormalized at the scale $\mu$ 
are given by:\footnote{The convention for the dimensional
parameter $\epsilon$ used here is: $d=4-2\epsilon$.} 
\bea
e_{V,A}^R(\mu )  
&=&
e_{V,A} -\frac{1}{2}(e_1+e_2) 
+\frac{3}{2}\left[\frac{1}{\epsilon}-\gamma_E+{\rm ln}(4\pi) 
+1  \right] 
+ 3 \;  {\rm ln}\left(\frac{\mu}{m_N}\right)
\, .
\label{eq;eJR}
\eea 
This renormalization is adequate
to remove all the infinities 
associated with virtual photons which  
we encounter in this calculation. 
The differential neutron decay rate including 
the radiative corrections 
and $1/m_N$ corrections
is found to be: 
\bea 
\frac{d\Gamma}{dE_e d\Omega_{\hat{p}_e} d\Omega_{\hat{p}_\nu} } 
=
\frac{(G_FV_{ud})^2}{(2\pi)^5}
\frac{ F(Z,E_e) \; |\vec{p}_e| \; E_\nu}{ m_n \; 
[E_p+E_\nu+E_e\; (\vec{\beta}\cdot\hat{p}_\nu)] }
|M|^2, 
\label{eq;theresult} 
\eea 
where
we have retained the relativistic expression 
for the phase factor, and
\bea 
\lefteqn{ |M|^2 
= m_nm_pE_eE_\nu 
\left( 1 +\frac{\alpha}{2\pi} \; e_V^R \right)
\left( 1 +\frac{\alpha}{2\pi} \; \delta_\alpha^{(1)} \right)
 }
\nnb \\ && \times C_0(E_e) (1+3 \gat^2) 
\left\{
1+ \left( 1 +\frac{\alpha}{2\pi} \; \delta_\alpha^{(2)} \right) 
C_1(E_e) \vec{\beta}\cdot\hat{p}_\nu
\right. \nnb \\ && \left.
+ \left( 1 +\frac{\alpha}{2\pi} \; \delta_\alpha^{(2)} \right)
[C_2(E_e)+C_3(E_e) \vec{\beta}\cdot\hat{p}_\nu ]
\hat{n}\cdot\vec{\beta}
+[C_4(E_e)+C_5(E_e) \vec{\beta}\cdot\hat{p}_\nu ]
\hat{n}\cdot\hat{p}_\nu
\right\} . 
\label{eq:matrel7}
\eea 
The explanation of the quantities appearing in this 
expression will be given below.  
We remark that, in order to arrive at this factored form, 
we have freely exploited the fact that terms of order 
$(\alpha/2 \pi)^2, (\alpha /2 \pi)(Q/ m_N)$ and 
$(Q/m_N)^2$ can be ignored to the order of accuracy of our 
concern.  

In Eq.~(\ref{eq;theresult}) the Coulomb part of the
radiative correction has been extracted as an overall factor 
and incorporated into the usual Fermi function
$F(Z,E_e) \simeq 1+(\alpha/2\pi) \delta_\alpha^{(Coul)}
 = 1+\alpha\pi/\beta$, for $Z=1$.  
In Eq.~(\ref{eq:matrel7}) 
the finite LEC, $e_V^R$, featuring in the factor 
$(1 +\frac{\alpha}{2\pi} \; e_V^R )$ 
subsumes those short-range radiative corrections 
to the Fermi constant $G_F^2$ which
have been integrated out in arriving at
our effective Lagrangian. 
This point will be further discussed in the final section.
The axial coupling constant, $g_A$, 
which has been renormalized by pion loops, 
is multiplied by short-range radiative corrections
involving the finite LEC $e_A^R$ as well as $e_V^R$. 
For convenience, and to simplify the results, we incorporate
this radiative correction to $g_A$
into $\gat$ defined by 
\bea 
\gat &=& g_A 
\left[
1+\frac{\alpha}{4\pi} ( e_A^R-e_V^R) 
\right] \, ,
\label{eq;gatilde}
\eea 
and this $\gat$ has been used in Eq.~(\ref{eq:matrel7}). 
Recall that $g_A$
corresponds to the physical value, 
with all short-range radiative corrections removed.

In Eq.~(\ref{eq:matrel7}), 
$\delta_\alpha^{(1)}$ represents 
the model-independent
radiative correction to $G_F$, which depends 
only on the kinematics of the electron, 
while $\delta_\alpha^{(2)}$ 
gives the model-independent radiative
corrections to the coefficients
of the angular correlation terms, 
$\vec{\beta}\cdot\hat{p}_\nu$ and 
$\hat{n}\cdot\vec{\beta}$. 
The explicit expressions for 
$\delta_\alpha^{(1)}$ and $\delta_\alpha^{(2)}$ are: 
\bea
\delta_\alpha^{(1)} &=& 
3\,{\rm ln}\left(\frac{m_N}{m_e}\right)
+ \frac{1}{2}
+ \frac{1+\beta^2}{\beta} {\rm ln}\left(\frac{1+\beta}{1-\beta}\right)
- \frac{1}{\beta}{\rm ln}^2\left(\frac{1+\beta}{1-\beta}\right)
+ \frac4\beta L\left(\frac{2\beta}{1+\beta}\right)
\nnb \\ &&
+ 4 \left[\frac{1}{2\beta}{\rm ln}\left(\frac{1+\beta}{1-\beta}\right)
-1\right]
\left[{\rm ln}\left(\frac{2(E_e^{max}-E_e)}{m_e}\right)
+ \frac{1}{3} \left(\frac{E_e^{max}-E_e}{E_e}\right)
-\frac{3}{2} 
\right] 
\label{eq:delta1}
\nnb \\ &&
+ \left(\frac{E_e^{max}-E_e}{E_e}\right)^2 \frac{1}{12\beta}
{\rm ln}\left(\frac{1+\beta}{1-\beta}\right) \, ,
\\
\delta_\alpha^{(2)} &=&
\frac{1-\beta^2}{\beta}{\rm ln}\left(\frac{1+\beta}{1-\beta}\right)
+\left(\frac{E_e^{max}-E_e}{E_e}\right)
\frac{4(1-\beta^2)}{3\beta^2}
\left[\frac{1}{2\beta}{\rm ln}\left(\frac{1+\beta}{1-\beta}\right)-1
\right]
\nnb \\ &&
+\left(\frac{E_e^{max}-E_e}{E_e}\right)^2
\frac{1}{6\beta^2}
\left[\frac{1-\beta^2}{2\beta}
{\rm ln}\left(\frac{1+\beta}{1-\beta}\right)-1
\right] \; . 
\label{eq:delta2}
\eea 
Here $E_e^{max}= (m_n^2-m_p^2+m_e^2)/2m_n$
is the maximum electron energy, 
and $L(z)$ is the Spence function defined by:
\begin{equation}
L(z)=\int^z_0 \frac{dt}{t}\ln (1-t).
\end{equation}
The factor $C_0(E_e)$ contains the recoil corrections to the overall
rate. 
It is given by 
\bea
C_0(E_e) =  1+\frac{1}{m_N (1+3\gat^2)} \left\{
 (\gat^2-2 \mu_V \gat+1)E_e^{max} -
\frac{m_e^2}{E_e}(1+\gat^2)+2 \mu_V \gat (\beta^2+1)E_e 
\right\}\, ,
\eea
where we have used $E_\nu=E_e^{max}-E_e +{\cal O}(1/m_N)$.  
The other coefficients  $C_i(E_e)$ ($i = 1,2,\ldots, 5$)
are given by 
\bea
C_1(E_e)&=&\tilde{a} \left\{1+\frac{1}{m_N}\left[
\frac{(\gat^2+2 \mu_V \gat+1)}{1+3 \gat^2} \frac{m_e^2}{E_e} +
\frac{(\gat^2+1)[8 \mu_V \gat E_e-4E_e^{max} \gat (\gat+\mu_V)]}
{(\gat^2-1)(1+3 \gat^2)}\right] \right\}\, ,  
\\
C_2(E_e)&=&\tilde{A} \left\{1+\frac{1}{m_N}\left[
\frac{(\gat^2-1)(\gat+\mu_V)}{2 \gat (1+3 \gat^2)}(E_e^{max}- E_e) +
\frac{E_e (\mu_V-1)}{\gat-1} -
\beta^2 E_e \frac{\gat^2+2 \gat \mu_V +1}{1+3 \gat^2}\right] 
\right\}\, ,   
\nnb \\ && \\
C_3(E_e)&=&\tilde{A} \frac{E_e (\gat-\mu_V)}{2 m_N \gat} \, ,    
\\
C_4(E_e)&=&\tilde{B} \left\{1+\frac{1}{m_N }\left[ 
\frac{E_e \beta^2 (\gat^2-1)(\gat-\mu_V)}{2 \gat (1+3 \gat^2)}+
\frac{(\gat+\mu_V)(\gat-1)^2}{(\gat+1)(1+3
\gat^2)}(E_e-E_e^{max})\right] 
\right\} \, , 
\\
C_5(E_e)&=&\tilde{B}\frac{(\gat+\mu_V)}{2 m_N \gat}(E_e^{max}-E_e)  \, ,
\eea
where $\tilde{a},\tilde{A},\tilde{B}$ are given by
Eq.~(\ref{eq;gA}) with the substitution $g_A \rightarrow \gat$. 
It is to be noted that Eq.~(\ref{eq:matrel7}) 
exhibits angular dependences that
are missing in Eq.~(\ref{eq;aAB}). 
These extra angular dependences
arise from the NLO contributions
that have been included in the $1/m_N$ corrections
(which leads to  Eq.~(\ref{eq:matrel7}))
but ignored in the LO evaluation 
(which leads to Eq.~(\ref{eq;aAB})). 
It has been a common practice to approximate 
the overall kinematic factor in Eq.~(\ref{eq;theresult})
by applying an expansion in $1/m_N$. 
If convenient, one could use the following approximation:
\bea
\frac{m_p E_\nu^2}{(E_p+E_\nu+E_e \vec{\beta}\cdot\hat{p}_\nu)} \simeq
(E_e^{max}-E_e)^2 \left[1+\frac{1}{m_N}(3 E_e-E_e^{max}-
3 E_e\vec{\beta}\cdot\hat{p}_\nu) \right], 
\label{eq;expphsp2}
\eea
where we have used 
\begin{equation}
E_{\nu}\simeq
(E^{max}_e-E_e)\left[
1+\frac{E_e}{m_N}
(1- \vec{\beta}\cdot\hat{p}_{\nu} )
\right]\, . 
\label{eq:Enuexpan}
\end{equation}
The angular dependence appearing in Eq.~(\ref{eq;expphsp2})
needs to be considered simultaneously with 
the angular dependences contained in  
Eq.~(\ref{eq:matrel7}).  

The model independent radiative correction 
$\delta_\alpha^{(1)}$ in Eq.~(\ref{eq:delta1})
agrees with that obtained by Sirlin~\cite{sirlin},
while 
$\delta_\alpha^{(2)}$ in Eq.~(\ref{eq:delta2}) 
also agrees with the result reported by 
Garcia and Maya~\cite{garcia}.  
We note that recoil corrections 
have also been calculated in the literature
using the conventional methods.
For instance, Wilkinson~\cite{wilkinson} evaluated
corrections to the decay rate and the 
correlation coefficient $A$, 
and Bilen'kii {\it et al.}~\cite{bilenky}
computed corrections to the decay rate and 
the correlation coefficient $a$.
Furthermore, Holstein~\cite{holstein} considered
recoil corrections to all the observables
for general nuclear beta-decays.
Our results for the recoil corrections
agree with those found in these previous studies.

\vskip 3mm \noindent
{\bf 4. Discussion and conclusions} 

As mentioned in the introduction,
a prime issue in the studies of neutron beta-decay
is to deduce the precise value of $V_{ud}$ 
from the experimental data.
Another issue is the extraction of the value
of $g_A$  from the data.
We shall discuss here the significance 
of our present calculation
in connection with these two issues.

To obtain the actual numerical values of 
$V_{ud}$ and $g_A$
we need to know the values of
the LEC's, $e_V^R$ and $e_A^R$, 
pertaining to the lepton-current nucleon-current vertex. 
These LEC's 
parameterize short-distance physics not explicitly 
included in the effective Lagrangian, ${\cal L}_\beta$,
and they need to be determined empirically using 
appropriate observables.
This is an important line of studies for the future.
Here, instead, we discuss simple 
order-of-magnitude estimates of the LEC $e_V^R$, 
which is the most important LEC in neutron beta-decay. 
Based on the general estimation of 
a photon loop diagram,
one may expect the natural scale for this parameter 
to be of the order of 
$(\alpha /2\pi)e_V^R \sim 2\times 10^{-2}$,
with $e_V^R\sim \ln(m_e/\Lambda_\chi)$. 
To obtain 
another rough estimate of $e_V^R$ 
we may compare our result 
for the neutron decay rate obtained
from Eq.~(\ref{eq;theresult}) 
with Eq.~(6) of Marciano and Sirlin~\cite{marciano-sirlin}.
Thus we introduce the premise
\be
e_V^R \simeq 
-\frac{5}{4}
-4\, {\rm ln}\left(\frac{m_W}{m_Z}\right) 
+3\, {\rm ln}\left(\frac{m_W}{m_N}\right)
+ {\rm ln}\left(\frac{m_W}{m_A}\right)
+ 2 C 
+ A_g  \, ,
\label{eq;eVRcomp}
\ee
where $m_W, m_Z$ are the masses of the W, Z bosons 
and $m_A$ is the axial mass scale.
As is customary, 
we define the Fermi constant $G_F$ of muon decay 
by absorbing the factor $1+(3\alpha/4 \pi){\rm ln}(m_W/m_Z)$ 
into $G_F$~\cite{sirlinnp}. 
The contribution 
${\rm ln}(m_W/m_Z)$ in Eq.~(\ref{eq;eVRcomp}) 
is actually the difference between the
contribution of the Z-box diagrams
\footnote{
The Z-box diagrams here refer to diagrams
like the one in Fig.~\ref{fig;diagrams} (f),
with the photon replaced by the Z boson;
see Fig. 3 in Ref. \cite{sirlinnp}.} 
in neutron beta-decay 
and the contribution of the Z-box diagrams
in muon decay.
In Eq.~(\ref{eq;eVRcomp}), 
the major contributions to 
the right-hand side
originate from the short-range virtual photon corrections 
to the Fermi transition
from the weak vector and axial-vector vertices.
The former gives the contribution,  $3\ln(m_W/m_N)$,
and the latter $\ln(m_W/m_A)$.
The $C$ in the expression
is the long-range model-dependent correction 
coming from the axial-current
and anomalous magnetic moments
of the nucleon,
and is proportional to 
$(\mu_S\, g_A)$ where $\mu_S$ is 
the isoscalar magnetic moment of
the nucleon. 
A value of $2C$ = 1.77 was found in 
Ref.~\cite{marciano-sirlin}.
In an HB$\chi$PT calculation, however, 
we have verified that a correction 
estimated from the diagrams of $C$ 
is of higher order $\propto \alpha /(2\pi)\; (\bar{Q}/m_N)^2$ 
and can be neglected (see section 2). 
Finally, the $A_g$ term, which includes 
a short-range strong-interaction correction,
is very small: 
$A_g \simeq -0.34$ \cite{marciano-sirlin}. 
In this connection,
it might be of interest to decompose, 
following Cirigliano {\it et al.}\cite{cirigliano03},
our $e_V^R$ into two parts:
$e_V^R=e_V^{SD}+\tilde{e}_V^R$.
The $e_V^{SD}$ term
describes the universal short-distance physics 
of electroweak theory discussed by Sirlin\cite{sirlinnp},  
while the $\tilde{e}_V^R$ term describes 
short-distance hadronic physics.
It is possible that the $A_g$ term
is associated with the $\tilde{e}_V^R$ term.
The above considerations lead to  
a rough estimate, $e_V^R\simeq 20$, {\it i.e.}, 
$[\alpha /(2\pi)] e_V^R \sim 4\times 10^{-2}$,
which is of a natural size as discussed above.
The above comparison also leads us to expect that
the dominant contribution to $e_V^R$ comes from 
the short-range electroweak corrections. 

The LEC $e_A^R$ enters only as a radiative
correction to $g_A$ in Eq.~(\ref{eq;gatilde}),
and therefore it may seem that there is no significant
motivation to remove the 
radiative correction $\frac{\alpha}{4\pi}(e_A^R-e_V^R)$ 
from $\gat$ defined in Eq.~(\ref{eq;gatilde})
and deduce the values of $g_A$. 
Indeed, if we limit ourselves to neutron beta-decay,
all the observables can be expressed using $\gat$
without referring to $g_A$.
However, since radiative corrections are specific
to individual processes,
there should be cases wherein the removal of 
$\frac{\alpha}{4\pi}(e_A^R-e_V^R)$
from $\gat$ has physical consequences and hence
$e_A^R$ does play a significant role.
A possible example is the
Goldberger-Treiman relation,
$g_A m_N = f_\pi\, g_{\pi N}$, where $g_{\pi N}$ is the 
pion-nucleon coupling constant.
To elaborate on this point,
it is useful to illustrate
processes which necessitate  
the introduction of the LEC, $e_A^R$.
To this end, 
we consider diagrams containing 
the exchange of a pion 
(pion-pole) plus a virtual photon. 
These diagrams involve three distinct
one-particle-irreducible vertex functions.
The first type is
a nucleon-nucleon-lepton-lepton four-point vertex
in which a virtual photon couples to both the nucleon 
and the leptonic currents.
This class of diagrams requires 
a counter term involving $e_A$
associated with $g_A$.
The second type is a lepton-lepton-pion 
three-point vertex
wherein a 
virtual photon only couples to the pion, the  
pion-lepton vertex or the lepton,
and this vertex is 
related to the pion decay constant $f_\pi$.
Some of the LEC's arising from this type of diagrams
can be found in the chiral Lagrangian considered by Knecht
{\it et al.}~\cite{knecht}.
These LEC's are also related to the ``inner" 
radiative corrections
calculated for pion beta-decay, 
see {\it e.g.}~\cite{marciano}.
The third type is a nucleon-nucleon-pion vertex 
in which a virtual
photon only couples to 
the pion, the pion-nucleon vertex or the nucleon,
and this vertex is
related to $g_{\pi N}$.
The corresponding LEC's are the $g_i$'s
appearing in M\"uller and Mei\ss ner's work~\cite{muller}. 
To our knowledge, however,
no systematic HB$\chi$PT study
of the Goldberger-Treiman relation
including the radiative corrections 
associated with each of the vertices
has been done so far.
In fact the radiative correction $e_A^R$ really has not been fully
studied yet in the standard approach. Instead it has usually been
assumed that $e_A^R \simeq e_V^R$, which makes the radiative
correction to $g_A$ small.
Such radiative corrections could contribute to the evaluation of the
Goldberger-Treiman discrepancy, but there is clearly not yet enough
information to determine whether they turn out to be significant in
comparison with the chiral symmetry breaking term.

As discussed in section 2, we have not included in our work 
radiative corrections involving the NLO vertex
or pion loop diagrams.
The former should be suppressed at least by a factor of
$\mu_V \bar{Q}/(2 m_N) \simeq 2 \times 10^{-3}$,
and the latter by a factor of
$(m_\pi/\Lambda_\chi)^2 \simeq 2 \times 10^{-2}$ relative to the
leading radiative corrections. 
Also omitted from our work
are the isospin breaking effects, 
which are naturally incorporated 
in the 
N$^2$LO heavy-baryon chiral 
Lagrangian~\cite{bkm95} 
not explicitly written in this paper.  
Recently, Kaiser \cite{kaiser} studied
isospin violation corrections to $G_F V_{ud} $ using  
HB$\chi$PT and found that the isospin
breaking corrections are of the order of 10$^{-5}$. 
To the accuracy of our present concern, we can safely 
neglect the isospin violation corrections.
    
We now summarize.  
Using the 
effective field theory for neutron beta-decay,
we have calculated the decay rate of the neutron 
and the angular correlation coefficients 
including recoil corrections and
radiative corrections of order $\alpha$. 
We have included 
all non-radiative terms through N$^2$LO 
except those which are negligible
because of the extremely small value of $\bar{Q}$ 
for neutron beta-decay.   
Our results reproduce the model-independent radiative
corrections and recoil corrections in the literature.  
The short-range radiative corrections of the earlier calculations
are replaced in our theory by 
the two finite radiative LEC's, 
$e^R_V$ and $e^R_A$,
where $e^R_V$ affects $G_F$ and
the difference, $e^R_V-e^R_A$,
affects $g_A$. 
Via comparison with the results of the existing model calculations,
we have argued that the value of $e_V^R$ is of a natural scale. 
An advantage of our EFT approach is the possibility of evaluating
higher order corrections in a systematic way, 
and the possibility to parameterize the strong interaction 
dependent contributions in terms of well-defined LEC's,
which can in principle be obtained from independent experiments. 
The next order corrections in 
the EFT for neutron beta-decay are 
estimated to be of the order 10$^{-5}$ or smaller.
Therefore, to the extent that
the LEC's involved in the present calculation
are of a ``natural" size (as discussed above),
we expect our expressions for the rate and 
the angular correlation coefficients to be 
accurate to better than $10^{-3}$. 

\vskip 3mm \noindent
{\bf Acknowledgments}
 
We thank T.-S. Park for discussions 
at the early stage of this work. 
SA thanks C.-P. Liu for discussions.
This work is supported 
in part by the Natural Sciences 
and Engineering Research Council of Canada,
by the Japan Society for the Promotion of Science, 
Grant No. 15540275,
by the United States National Science Foundation,
Grant No. PHY-0140214,
and by the United States Department of Energy,
Grant No. DE-FG02-03ER46043.


\begin{thebibliography}{99}

\bibitem{Vud}
J.~C. Hardy, I.~S. Towner, V.~T. Koslowsky,
E. Hagberg, and H. Schmeing,
Nucl. Phys. A {\bf 509} (1990) 429.

\bibitem{cirigliano03} 
V. Cirigliano, M. Knecht, H. Neufeld, H. Rupertsberger, and P. Talavera, 
Eur. Phys. J. C{\bf 23} (2002) 121; 
V. Cirigliano, M. Knecht, H. Neufeld, and H. Pichl,
Eur. Phys. J. C{\bf 27} (2003) 255. 

\bibitem{sirlin}
A. Sirlin, Phys.\ Rev. {\bf 164} (1967) 1767.

\bibitem{sirlinrmp}
A. Sirlin, Rev.\ Mod.\ Phys. {\bf 50} (1978) 573.

\bibitem{marciano-sirlin}
W.~J. Marciano and A. Sirlin, 
Phys.\ Rev.\ Lett. {\bf 56} (1986) 22.

\bibitem{bunatian}
G.~C.~Bunatain, arXiv:hep-ph/0311350.

\bibitem{weinberg}
S. Weinberg, Physica A {\bf 96} (1979) 327.

\bibitem{bkm95} 
V. Bernard, N. Kaiser, and U.-G. Mei\ss ner, 
Int. J. Mod. Phys. {\bf E4} (1995) 193. 

\bibitem{scherer}
S. Scherer, 
In ``Advances in Nuclear Physics'', Vol. 27,
edited by J.~W. Negele and E. Vogt. 

\bibitem{muller}
G. M\"uller and U.-G. Mei\ss ner,
Nucl.\ Phys.\ B {\bf 556} (1999) 265.

\bibitem{fettes2}
N. Fettes and U.-G. Mei\ss ner, 
Nucl.\ Phys.\ A, {\bf 693} (2001) 693.

\bibitem{fear97} 
H.~W. Fearing, R. Lewis, N. Mobed, and S. Scherer, 
Phys.\ Rev.\ D {\bf 56} (1997) 1783.

\bibitem{bern98} 
V. Bernard, H.~W. Fearing, T.~R. Hemmert, and U.-G. Mei\ss ner, 
Nucl.\ Phys.\ A {\bf 635} (1998) 121.

\bibitem{ando00} 
S. Ando, F. Myhrer, and K. Kubodera,
Phys.\ Rev.\ C {\bf 63} (2000) 015203.

\bibitem{bern01} 
V. Bernard, T.~R. Hemmert, and U.-G. Mei\ss ner,
Nucl.\ Phys.\ A {\bf 686} (2001) 290.

\bibitem{knecht} 
M. Knecht, H. Neufeld, H. Rupertsberger, and P. Talavera, 
Eur.\ Phys.\ J.\ C {\bf 12} (2000) 469.

\bibitem{bgmzRNBD}
V. Bernard, S. Gardner, U.-G. Mei\ss ner, C. Zhang,
arXiv:hep-ph/0403241, to appear in Phys. Lett. B.

\bibitem{jtw57} 
J.~D. Jackson, S.~B. Treiman, and H.~W. Wyld, 
Phys.\ Rev.\ {\bf 106} (1957) 517.  

\bibitem{ct-pr67}
C.~G. Callan and S.~B. Treiman, 
Phys. Rev. {\bf 162} (1967) 1494.

\bibitem{garcia}
A. Garc\'{i}a and M. Maya,
Phys.\ Rev.\ D {\bf 17} (1978) 1376.

\bibitem{wilkinson}
D.~E. Wilkinson, 
Nucl.\ Phys.\ A {\bf 377} (1982) 474.

\bibitem{bilenky}
S.~M. Bilen'kii, R.~M. Ryndin, Ya.~A. Saoridinski\v{i}, and Ho Tso-Hsiu,
Sovi. Phys. JETP {\bf 37} (1960) 1241.

\bibitem{holstein}
B.~R. Holstein, 
Rev. Mod. Phys. {\bf 46} (1974) 789;
Erratum ibid {\bf 48} (1976) 673.

\bibitem{sirlinnp}
A. Sirlin, Nucl.\ Phys.\ B {\bf 71} (1974) 29.

\bibitem{marciano}
W.~J.~Marciano and A.~Sirlin, 
Phys.\ Rev.\ Lett. {\bf 71} (1993) 3629.

\bibitem{kaiser}
N. Kaiser, Phys.\ Rev.\ C {\bf 64} (2001) 028201.

\end{thebibliography}
\end{document}